\documentclass[a4paper,runningheads]{llncs}

\pdfoutput=1
\usepackage{hyperref}
\hypersetup{
	pdftitle={Author Name Disambiguation by Using Deep Neural Network}, 
	pdfauthor={Hung Nghiep Tran, Tin Huynh, Tien Do}, 
	pdfsubject={Computer Science}, 
	pdfkeywords={Artificial Intelligence, Machine Learning, Digital Library, Deep Learning, Bibliographic Databases, Feature Learning, Deep Neural Network}, 
	pdffitwindow=false,     
	pdfstartview={FitH},    
	pdfcreator={Hung Nghiep Tran}, 
	pdfproducer={Hung Nghiep Tran} 
}

\usepackage{mathtools}
\usepackage{cite}
\usepackage{amsmath, amssymb}
\usepackage{graphicx}
\usepackage{array}
\usepackage{tabularx}
\usepackage{multirow}
\usepackage{hhline}
\usepackage{verbatim} 
\usepackage{url}

\makeatletter
\def\hlinewd#1{%
\noalign{\ifnum0=`}\fi\hrule \@height #1 %
\futurelet\reserved@a\@xhline}
\makeatother

\urldef{\mailsa}\path|{alfred.hofmann, ursula.barth, ingrid.haas, frank.holzwarth,|
\urldef{\mailsb}\path|anna.kramer, leonie.kunz, christine.reiss, nicole.sator,|
\urldef{\mailsc}\path|erika.siebert-cole, peter.strasser, lncs}@springer.com|
\newcommand{\keywords}[1]{\par\addvspace\baselineskip
\noindent\keywordname\enspace\ignorespaces#1}

\newcolumntype{R}[1]{>{\raggedright\arraybackslash}p{#1}} 
\newcolumntype{C}[1]{>{\centering\arraybackslash}p{#1}} 
\newcolumntype{L}[1]{>{\raggedleft\arraybackslash}p{#1}} 

\newcolumntype{+}{>{\global\let\currentrowstyle\relax}}
\newcolumntype{^}{>{\currentrowstyle}}
\newcommand{\rowstyle}[1]{\gdef\currentrowstyle{#1}%
#1\ignorespaces
}

\DeclareMathSizes{10}{10}{8}{8}   
\DeclareMathSizes{11}{11}{9}{9}   
\DeclareMathSizes{12}{12}{10}{10}  

\begin{document}

\mainmatter  

\title{ Author Name Disambiguation \\by Using Deep Neural Network}

\titlerunning{Author Name Disambiguation by Using Deep Neural Network}

%
%

\author{Hung Nghiep Tran, Tin Huynh, Tien Do}

\authorrunning{Hung Nghiep Tran et al.}

\institute{
University of Information Technology - Vietnam, \\
Km 20, Hanoi Highway, Linh Trung Ward, Thu Duc District, HCMC, Vietnam. \\
\email{\{nghiepth, tinhn, tiendv\}@uit.edu.vn}
}


%
%


\maketitle

\begin{abstract}
Author name ambiguity is one of the problems that decrease the quality and reliability of information retrieved from digital libraries. Existing methods have tried to solve this problem by predefining a feature set based on expert's knowledge for a specific dataset. In this paper, we propose a new approach which uses deep neural network to learn features automatically for solving author name ambiguity. Additionally, we propose the general system architecture for author name disambiguation on any dataset. We evaluate the proposed method on a dataset containing Vietnamese author names. The results show that this method significantly outperforms other methods that use predefined feature set. The proposed method achieves 99.31\% in terms of accuracy. Prediction error rate decreases from 1.83\% to 0.69\%, i.e., it decreases by 1.14\%, or 62.3\% relatively compared with other methods that use predefined feature set (Table \ref{tab:Evaluation}).

\keywords{Digital Library, Bibliographic Data, Author Name Disambiguation, Machine Learning, Feature Learning, Deep Neural Network.}
\end{abstract}

\section{Introduction}
Author name ambiguity is a problem that occurs when a set of publication records contains ambiguous author names, i.e., the same author may appear under distinct names (synonymy), or distinct authors may have similar names (polysemy)  \cite{Ferreira:2012:BSA:2350036.2350040}. This problem decreases the quality and reliability of information retrieved from digital libraries such as the impact of authors, the impact of organizations, etc. Therefore, author name disambiguation is a critical task in digital libraries.

There are two approaches to author name disambiguation: (1) grouping publication records of a same author by finding some similarity among them (author grouping methods) or (2) directly assigning them to their respective authors (author assignment methods) \cite{Ferreira:2012:BSA:2350036.2350040}. Both of them try to create, select and combine features based on the similarity of attributes (author names, keywords, etc.) by using some measures such as Jaccard, Jaro, etc., or some heuristics. However, most of those works are done manually based on experts' knowledge. Each predefined feature set could perform very well on a specific dataset that experts originally dealt with, but it could perform poorly on other datasets. To solve this problem, a method to learn features automatically from data is necessary.

Neural networks, which have many layers, are called deep neural networks (DNN). Recent researches \cite{NIPS2012_0534, DBLP:conf/cvpr/CiresanMS12, DBLP:journals/corr/abs-1301-3605} have shown their strong ability in feature learning in many tasks. Internal features learned by the DNN are relatively stable for variants in data if the training data are sufficiently representative \cite{DBLP:journals/corr/abs-1301-3605}. This helps dealing with citation errors\footnote{Citation errors: errors in citation data which are sometimes impossible to detect.}, which is an open challenge pointed out by Ferreira et al. \cite{Ferreira:2012:BSA:2350036.2350040}. Moreover, using neural network has the advantage that it would build a general model. This model could disambiguate author name incrementally when new publication records are incorporated into the dataset.

In this paper, we propose a new approach which uses deep neural network to learn features automatically for solving author name ambiguity. Additionally, we propose the general system architecture for author name disambiguation on any dataset. This system computes a representative for a dataset, and then uses a combination of many DNNs to learn features and disambiguate author names.

The remainder of the paper is organized as follows. Section 2 briefly presents related researches on author name disambiguation and feature learning using DNN. Section 3 will describe the proposed method and system architecture. In section 4, we will present the experiments, evaluation results and discussions. Finally, we conclude the paper and suggest future works in section 5.

\section{Related Work}
Ferreira et al. \cite{Ferreira:2012:BSA:2350036.2350040} did a brief survey of author name disambiguation methods. According to their survey, existing methods have tried to create, select and combine features based on the similarity of attributes by using some string-matching measures or some specific heuristic, such as the number of coauthor names in common, etc.

Bhattacharya and Getoor \cite{Bhattacharya:2007:CER:1217299.1217304} proposed a combined similarity function defined on attributes and relational information. The method obtained a high F1 score around 0.99 in the CiteSeer collection, lower in the arXiv collection and only around 0.81 in the BioBase collection.

In another research, Torvik et al. \cite{Torvik:2005:PSM:1059557.1059561} used a feature set resulting from the comparison between the common citation attributes along with medical subject headings, language, and affiliation of two references in MEDLINE dataset. In a subsequent work \cite{Torvik:2009:AND:1552303.1552304}, Torvik and Smalheiser incorporated some features into their method to achieve better result.

In our previous research \cite{Huynh:2013:VAN:2450480.2450508}, we predefined a feature set to learn a similarity function specifically for Vietnamese author dataset, one of the most difficult case, and obtained around 0.98 of accuracy.

In those researches, the central task is predefining a feature set for a specific dataset. A good feature set will help improving accuracy on a specific dataset, but it need to be recalibrated for a new dataset. In this research, we aim at learning features automatically.

DNN could be regarded as feedforward neural network with more than one layer \cite{Rumelhart:1986:LIR:104279.104293}. Recently, many training and initialization schemes have been proposed in order to improve learning speed on such deep network, e.g., such as RBM \cite{Hinton:2006:FLA:1161603.1161605}, sparse auto-encoders \cite{DBLP:conf/nips/RanzatoBL07}, and normalized initialization \cite{DBLP:journals/jmlr/GlorotB10}. Deep learning using DNN has been a popular method for automatic feature learning in many tasks.

Ciresan et al. \cite{DBLP:conf/cvpr/CiresanMS12} was very successful in using a big DNN to learn features in image recognition. They built a deep convolution neural network and trained such network by simple online back-propagation. Their models greatly outperformed previous methods on many well-known datasets such as MNIST \footnote{\url{http://yann.lecun.com/exdb/mnist/}}, NORB \footnote{\url{http://www.cs.nyu.edu/~ylclab/data/norb-v1.0/}}, etc. without using complicated image pre-processing techniques.

Yu et al. \cite{DBLP:journals/corr/abs-1301-3605} used a simple deep feedforward neural network to learn features in speech recognition. They proved the model's ability to extract discriminative internal features that are robust to variants in data. Their model outperformed state-of-the-art systems based on GMMs or shallow networks without the need for explicit model adaptation or feature normalization.

However, to the best of our knowledge, there has not been any research that attempts to learn feature automatically by using DNN for author name disambiguation. Therefore, in this research, we will explore that approach.

\section{Our Approach}
In this section, we describe our proposed method and the general system architecture for author name disambiguation on any dataset. This method uses a combination of many DNNs to learn features from a data representative, which could be computed automatically for any dataset.

\subsection{Deep Neural Network}
DNN is a popular method for automatic feature learning \cite{Rumelhart:1986:LIR:104279.104293}. Some recent researches have successfully exploited feature learning using DNN to achieve state-of-the-art performance in many tasks \cite{DBLP:conf/cvpr/CiresanMS12, DBLP:journals/corr/abs-1301-3605, NIPS2012_0534}. There are many types of DNN. All of them are neural networks with many layers, but they are different in parameter initialization scheme, training algorithm, activation function, etc.

\begin{figure}
\centering
\includegraphics[width=0.45\linewidth]{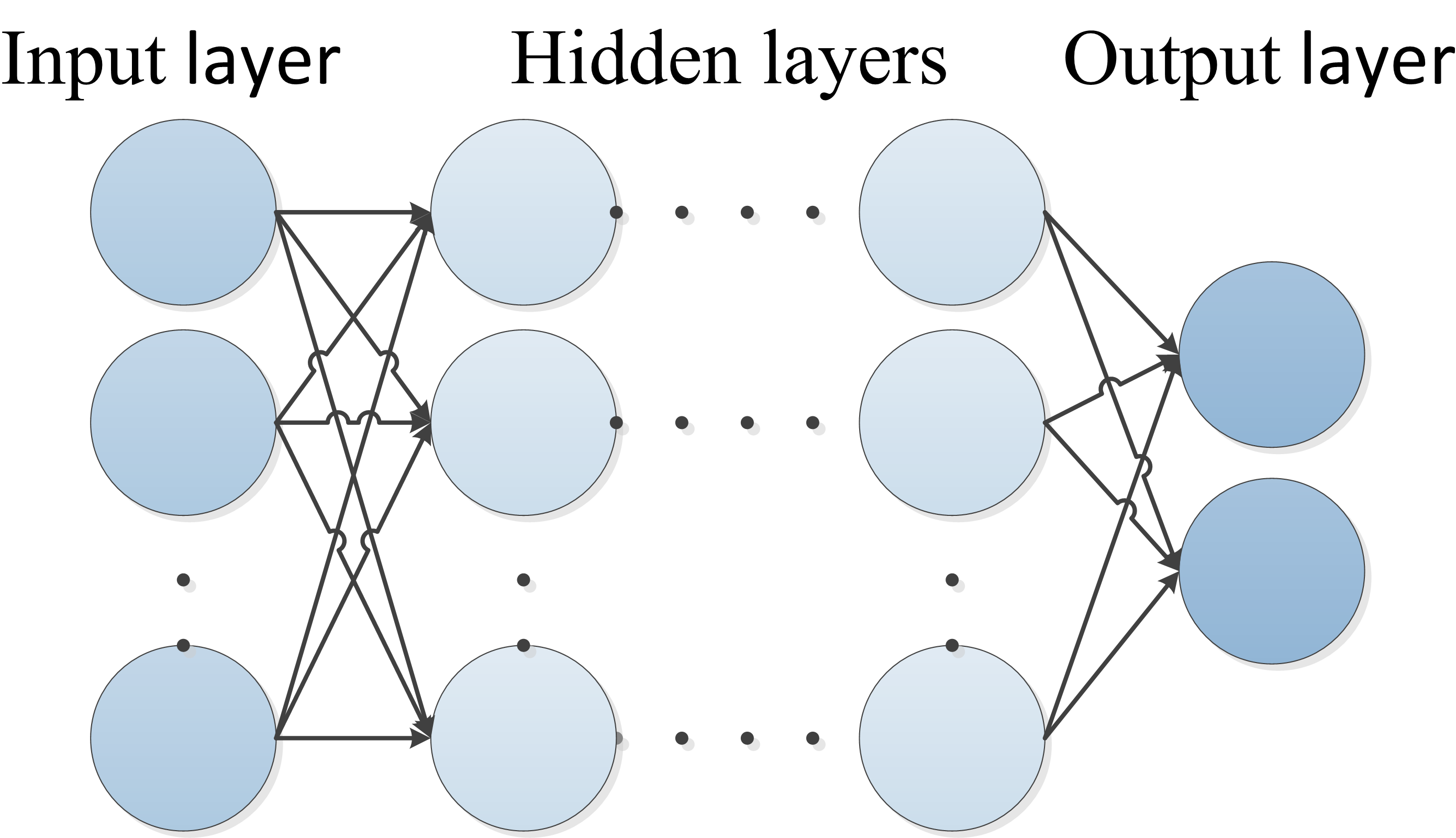}
\caption{Deep Neural Network Architecture}
\label{fig:DNNArchitecture}
\end{figure}

In this research, we use DNN with simple feedfoward architecture, a.k.a., multi-layer perceptron \cite{Rumelhart:1986:LIR:104279.104293}. Figure \ref{fig:DNNArchitecture} shows the general architecture of such network. The network has many layers stacked upon each other. Each neuron unit in each layer connects to every unit in the sequential layer. The number of units in the input layer corresponds with the number of basic features we use. Output layer contains two units which correspond with the case where two citations belong to the same author and otherwise, respectively.

If we denote the input and the ideal output of the DNN as x and y, respectively, a DNN can be interpreted as a directed graphical model that approximates the posterior probability $p_{y|x}(y = c|x)$ of a class $c$ given an input $x$. 

We consider a DNN with $L-1$ hidden layers and $1$ output layer, which is a stack of $L$ layers of log-linear models. Each hidden layer $l$ models the posterior probabilities of a hidden vector $h^l$ given the preceding layer's output $v^l$. If $h^l$ consists of $N^l$ hidden units, each denoted as $h_j^l$, the posterior probabilities, with sigmoid activation function for simplicity, can be expressed as 
\begin{equation}
\label{eq:InternalProbability}
p^l (h^l|v^l) = \prod\limits_{j=1}^{N^l} \frac{1} {1 +  e^{-z_j^l(v^l)}}, 0 < l < L
\end{equation}
where $z^l(v^l) = W^l\cdot v^l + a^l$, and $W^l$ and $a^l$ represent the weight matrix and bias vector in the $l-th$ layer, respectively.

Internal features will be learned in each hidden layer \cite{DBLP:journals/corr/abs-1301-3605}. Each hidden unit's output represents an internal feature and each hidden layer's output composes an internal feature vector. Starting with the basic feature input $v^1 = h^0 = x$, the output of each layer becomes the input of the next one, i.e., $v^{l+1} = h^l$. The latter layer will learn more sophisticated features.

In the final layer, the class posterior probabilities are computed as a multinomial distribution using softmax
\begin{equation}
\label{eq:OutputProbability}
p_{y|x}(y = c|x) = p^L(y = c|v^L) = \frac{e^{z_c^L(v^L)}} {\sum_{c'} e^{z_{c'}^L(v^L)}}
\end{equation} 

This type of DNN has the vanishing gradient issue when being trained with the traditional activation function $sigmoid(x)=1/(1+e^{-x})$ and back-propagation algorithm. To address this issue, we use the activation function $softsign(x)=x/(1+|x|)$ and the adaptive resilient backpropagation algorithm together with some training techniques \cite{DBLP:journals/jmlr/GlorotB10}.

DNN's ability in a specific case is affected by the network's structure and its parameters. Network parameters could be learned by training using some optimization algorithms. Network structure includes two hyperparamters: the number of hidden layers and the number of hidden units in each layer. The number of units should be equal among hidden layers so that information could flow effectively \cite{DBLP:journals/jmlr/GlorotB10}. In general, deeper and larger network will achieve better results. However, this makes training slower and is capable to yield overfitting. 

Those two hyperparameters are usually chosen based on experiments on the validation set \cite{DBLP:conf/cvpr/CiresanMS12, DBLP:journals/corr/abs-1301-3605}. In this research, we determine the optimum network size based on experiments using k-fold cross-validation. We begin with a small network, and then change the number of hidden layers and the number of hidden units respectively to create networks at larger sizes. The optimum network structure is the one with the highest average validation accuracy.

\subsection{Data Representative}
As we have shown in the previous subsection, DNN could learn internal features from data. In order to do this, data must be put into the input layer in a proper way. The input should be a good representative for data, i.e., it could describe details in data. We call that data representative the basic feature set.

There are many different ways to compute a data representative. One obvious way is to measure the similarity between all attributes of two publication records such as Author name, Affiliations, Coauthor, and Paper keyword, etc. using string-matching measures. We assume that the similarity between those attributes expresses how much two publications belong to the same author.

According to some surveys reviewing string-matching measures to identify the duplication \cite{Bilenko:2003:ANM:1137237.1137369, CohenRF03}, there are three types of measure: (1) \emph{Edit distance} such as Levenshtein, Monger-Elkan, Jaro, and Jaro-Winkler; (2) \emph{Token-based} such as Jaccard and TF/IDF and (3) \emph{Hybrid measures} such as Mogne-Elkan for comparing two long strings. We employ all three types of measure.

Each publication record has different attributes. In general case, we could apply computations to all available attributes, and use default value when they are unavailable. Therefore, how we build the data representative does not depend on a specific dataset.

In this research, we use these measures: Jaccard, Levenshtein, Jaro, Jaro-Winkler, Smith-Waterman, Mogne-Elkan. We apply them for these attributes: author name, co-author, affiliation, paper keyword, author interest keyword.

\subsection{System Architecture}
\begin{figure}
\centering
\includegraphics[width=0.7\linewidth]{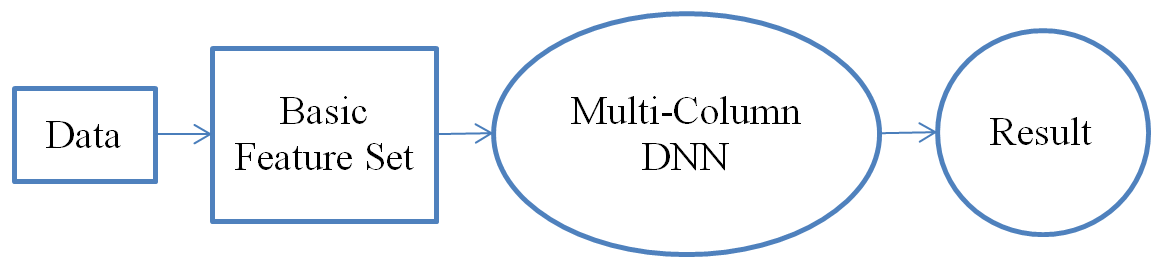}
\caption{System Architecture}
\label{fig:SystemArchitecture}
\end{figure}

In this subsection, we describe the general system architecture for author name disambiguation. The system could run on any dataset without expert's modifications. Figure \ref{fig:SystemArchitecture} shows the proposed system architecture. The system incorporates two components.

The first component takes the data and computes a data representative, i.e., a basic feature set $x$ to represent data. There could be many representatives for the same data, so this component could be implemented in many ways. In this research, we use string-matching measures to compute the data representative. The computations could be performed on any dataset automatically. 

The second component takes the basic feature set as its input, and then learns features in its hidden layers to disambiguate author names. The last layer of DNN computes the probabilities $p_{y|x}(y = c|x) = p^L(y = c|v^L)$ to determine whether two author instance names in a pair belong to the same author (when $p \geq 0.5$) or not (when $p<0.5$).

\begin{figure}
\centering
\includegraphics[width=0.6\linewidth]{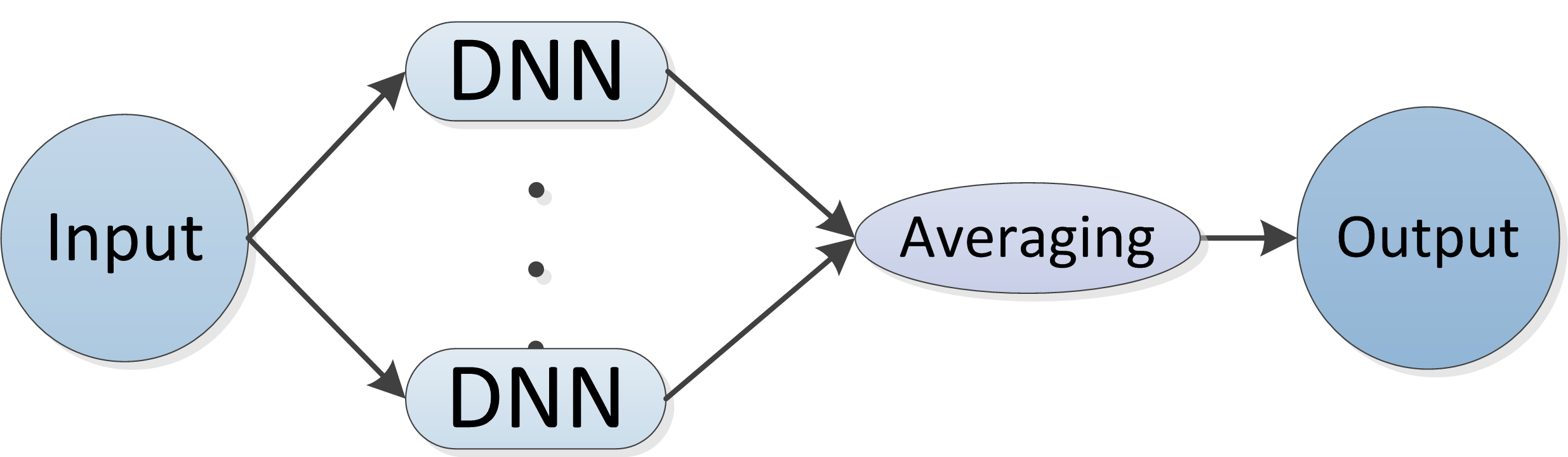}
\caption{Multi-Column Deep Neural Network}
\label{fig:MultiColumnDNN}
\end{figure}

In this system architecture, we use multi-column DNN technique, which is illustrated in figure \ref{fig:MultiColumnDNN}, to improve the generalization capabilities of the system \cite{DBLP:conf/cvpr/CiresanMS12}. This technique is similar to an ensemble method known as bootstrap aggregating or bagging.

Specifically, we will train $N$ DNNs simultaneously using data retrieved randomly from the training set in a manner similar to k-fold cross-validation. After training, we have $N$ distinct DNNs. In testing phase, we will apply all those DNNs simultaneously to each item in a separate test set. Then, we will take the final result by averaging results from those DNNs as
\begin{equation}
P_{y|x}(y = c|x) = \overline{p_{y|x}(y = c|x)} = \frac{\sum\limits_{n=1}^{N} p_{y|x}^n(y = c|x)}{N}
\end{equation}
where $p^n$ is the output of $n-th$ DNN.

\section{Experiment, Evaluation and Discussion}
We evaluate the effect of using automatic feature learning by DNNs for the author name disambiguation problem on Vietnamese author dataset (very ambiguous cases \cite{Ferreira:2012:BSA:2350036.2350040}), which we collected from online digital libraries. This section presents our experiment settings, evaluation results, and discussions.

\subsection{Dataset}

\begin{flushleft}
\textbf{Vietnamese author dataset:}
\end{flushleft}

In a previous work \cite{Huynh:2013:VAN:2450480.2450508}, we built a Vietnamese author dataset for checking author name ambiguity. Data was acquired from three online digital libraries that are ACM \footnote{\url{http://dl.acm.org}}, IEEE Xplore \footnote{\url{http://ieeexplore.ieee.org/Xplore/home.jsp}}, and MAS \footnote{\url{http://academic.research.microsoft.com/}} by querying their search engine using names of 10 Vietnamese authors. For these authors, there are many different instance names, e.g., author `Hoang Kiem' can have many instance names such as `Hoang Kiem', `Kiem Hoang', `Hoang Van Kiem', `Kiem Van Hoang', etc.

Query results are publications with different author instance names. We built the dataset by creating pairs of publications for each author. Based on our understanding about these authors, we manually labeled each pair with value 1 if ambiguous names in this pair actually were one person and value 0 otherwise.

In this research, we extend the dataset, so that, there are totally 30537 samples in the dataset. Table \ref{Tab:VietnameseData} shows the dataset details, the total number of pairs is counted without duplicated ones.

\begin{table}
\centering
    \caption{The Vietnamese author dataset}
    \label{Tab:VietnameseData}
    \begin{tabular}{>{\bfseries}+p{3.7cm}|^C{3.3cm}|^C{3.3cm}}
    \hlinewd{1.5pt}
    \rowstyle{\bfseries}
    Authors & Number of pairs with label 0 & Number of pairs with label 1 \\ \hline
    Cao Hoang Tru & 6522 & 409 \\
    Dinh Dien & 6750 & 384 \\
    Duong Anh Duc & 6757 & 406 \\
    Ha Quang Thuy & 6812 & 351 \\
    Ho Tu Bao & 6728 & 435 \\
    Hoang Kiem & 6753 & 410 \\
    Le Dinh Duy  & 6725 & 387 \\
    Le Hoai Bac & 6728 & 435 \\
    Nguyen Ngoc Thanh & 6869 & 294 \\
    Phan Thi Tuoi & 6728 & 435 \\
    \hline
    Total & 26591 (87.07\%) & 3946 (12.93\%)\\ 
    \hlinewd{1.5pt}
    \end{tabular}
\end{table}

\begin{flushleft}
\textbf{Dataset preparation:}
\end{flushleft}

For each pair of publication records, we compute all basic features. So each pair of publication records is represented as a basic feature vector with label 1 or 0. We use this vector as the input for the DNN.

From the original dataset, we hold out 20\% of the data as the test set for final performance evaluation. We use 5-fold cross-validation on the remaining 80\% of the data to tune hyperparameters and avoid overfitting. Each split dataset contains almost equal percentage of random samples of one particular class. Those samples are picked randomly at uniform distribution.

\subsection{Tuning Hyperparameters}
On Vietnamese author dataset, we experiment with 5-fold cross-validation to choose network size. We begin with the smallest network size of 1 hidden layer and 10 hidden units. We use this network as the baseline for hyperparameters tuning and achieve average validation accuracy of 95.35\%.

Then we increase the number of hidden layers and hidden units respectively and conduct experiments at many network sizes. Table \ref{tab:TuneSize} shows five network sizes (the number of hidden layers $\times$ the number of hidden units in each layer) with the highest accuracy. The network with 7 hidden layers and 50 hidden units in each layer achieves the highest average validation accuracy.

\begin{table}
\centering
    \caption{Five network sizes with the highest accuracy}
    \label{tab:TuneSize}
    \begin{tabular}{>{\centering\arraybackslash\hspace{0pt}}+p{3cm}|>{\centering\arraybackslash\hspace{0pt}}^p{6cm}}
    \hlinewd{1.5pt}
    \rowstyle{\bfseries}
    Network size & Average validation accuracy (\%) \\ \hline
    $7 \times 50$ & 99.35 \\
    $7 \times 75$ & 99.33 \\
    $6 \times 75$ & 99.28 \\
    $5 \times 75$ & 99.27 \\
    $6 \times 100$ & 99.25 \\
    \hlinewd{1.5pt}
    \end{tabular}
\end{table}

\subsection{Evaluation}
In our recent research \cite{Huynh:2013:VAN:2450480.2450508}, we proposed an approach based on a predefined feature set for Vietnamese author name and applied several classification models to that feature set. According to that research, k-NN, Random Forest, C4.5, SVM, and Naive Bayes, respectively, are the best suitable methods for the predefined feature set.

In this research, we compare the proposed method with those methods. The proposed method implements the system architecture that we have described. The DNNs use the hyperparameters that have been tuned. The other methods use the same settings and implementations as in our previous research \cite{Huynh:2013:VAN:2450480.2450508}.

Table \ref{tab:Evaluation} shows evaluation results in terms of accuracy and error on a separated test set. Results show that the proposed method significantly outperforms methods that use predefined feature set. The proposed method achieves 99.31\% in terms of accuracy. Whereas, the best method that uses predefined feature set achieves 98.17\% in terms of accuracy. Prediction error rate decreases from 1.83\% to 0.69\%, i.e., it decreases by 1.14\%, or 62.3\% relatively compared with other methods that use predefined feature set.

\begin{table}
\centering
    \caption{Evaluation results}
    \label{tab:Evaluation}
    \begin{tabular}{>{\bfseries}+p{4cm}|^p{2.3cm}|>{\centering\arraybackslash\hspace{0pt}}^p{2.3cm}|>{\centering\arraybackslash\hspace{0pt}}^p{2.3cm}}
    \hlinewd{1.5pt}
    \rowstyle{\bfseries}
    Feature set & Method & Accuracy (\%) & Error (\%) \\ \hline
    \multirow{5}{*}{Predefined} & k-NN & 98.17 & 1.83 \\
    \hhline{~---}
    & Random Forest & 98.13 & 1.87 \\
    \hhline{~---}
	& C4.5 & 98.02 & 1.98 \\
    \hhline{~---}
    & SVM & 97.45 & 2.55 \\
    \hhline{~---}
    & Naive Bayes & 96.68 & 3.32 \\ 
    \hline
    Learned Automatically & DNN & \textbf{99.31} & \textbf{0.69} \\
    \hlinewd{1.5pt}
    \end{tabular}
\end{table}

\subsection{Discussion}
Evaluation results clearly show benefits of learning features compared with predefining features in terms of accuracy. Moreover, automatic feature learning does not require expert's knowledge on specific dataset. DNN has been used to learn features successfully. However, due to its high capability to learn complex features, it is prone to overfitting.

In this research, we use many techniques to reduce overfitting. We extend Vietnamese author dataset. We use k-fold cross-validation for hyperparameter tuning and early stopping. Moreover, our system architecture uses multi-column technique to have a lower variance result.

The type of DNN we use has the vanishing gradient issue when being trained with the traditional sigmoid activation function and back-propagation algorithm. One solution is choosing a good activation function \cite{DBLP:journals/jmlr/GlorotB10}. The hyperbolic tangent function $tanh(x) = (1-e^{-2x})/(1+e^{-2x})$ is better than the traditional sigmoid function thanks to its zero mean. Whereas, the softsign function $softsign(x) = x/(1+|x|)$ is better than $tanh(x)$ thanks to its smoother asymptotic behavior. The rectifier function $max(x, 0)$ is one of the best activation functions \cite{DBLP:journals/jmlr/GlorotB10, DBLP:journals/jmlr/GlorotBB11}.

The current DNN model is supervised, therefore, it needs labeled data, which is usually not easy to obtain. However, there are some techniques to pre-train DNN using unlabeled data, which are special kind of weight initialization methods, to improve performance much in terms of training time and accuracy, especially in case lack of labeled data.

The proposed method is prospective when data is integrated from heterogeneous sources, because in such case, it is difficult to predefine a feature set. 

On the other hand, automatic feature learning using DNN has shown its ability in many complex tasks such as image recognition, where this method could recognize object just by using raw pixels \cite{DBLP:conf/cvpr/CiresanMS12}. Therefore, it is rational to think of creating such `pixels' in author name disambiguation by encoding bibliographic data and use those `pixels' as the basic data representative in the DNN.

\section{Conclusion and Future Work}
In this paper, we have proposed a new approach which uses deep neural network to learn features automatically for solving author name ambiguity. Additionally, we have proposed the general system architecture for author name disambiguation on any dataset.

We have evaluated the proposed method on a Vietnamese author dataset. The results show that this method significantly outperforms other methods that use predefined feature set. The proposed method achieves 99.31\% in terms of accuracy. Prediction error rate decreases from 1.83\% to 0.69\%, i.e., it decreases by 1.14\%, or 62.3\% relatively compared with other methods that use predefined feature set.

The proposed method could be extended to solve some open challenges such as the lack of labeled training data, incremental and new-author disambiguation.

In the future, we will benchmark the proposed method on other datasets. We will also experiment with other activation functions and unsupervised pre-training methods on encoded bibliographic data.\\\\
\textbf{Acknowledgments}. This research is funded by University of Information Technology, VNU-HCMC under grant number C2012-07.

\bibliographystyle{splncs03}

\end{document}